# Three-dimensional study of grain scale tensile twinning activity in Mg: A combination of microstructure characterization and mechanical modeling

Xun Zeng[a], Chuanlai Liu[b, *], Chaoyu Zhao[c], Jie Dong[c], Franz Roters[b], Dikai Guan[a, *]

[a] Department of Mechanical Engineering, University of Southampton, Southampton SO17 1BJ, UK

[b] Max-Planck-Institut für Eisenforschung GmbH, Max-Planck-Str. 1, Düsseldorf 40237, Germany

[c] National Engineering Research Center of Light Alloy Net Forming and State Key Laboratory of Metal Matrix Composite, School of Materials Science and Engineering, Shanghai Jiao Tong University, Shanghai 200240, China

[*]Corresponding authors

Email address: dikai.guan@soton.ac.uk (Dikai Guan) c.liu@mpie.de (Chuanlai Liu)

**Abstract:** Tensile twinning is a main deformation mode in hexagonal close packed structure metals, so it is important to comprehensively understand twinning mechanisms which are not fully disclosed using 2D or small volume 3D characterization techniques. A large area 3D electron backscatter diffraction (EBSD) measurement and crystal plasticity modeling were carried out to investigate the tensile twinning behaviors in a Mg alloy. The results showed that tensile twinning activity was underestimated using conventional 2D EBSD scans. When compressed to yield point, the examined twin frequency with 2D was lower than that using 3D EBSD. The effects of Schmid factor (SF) on twinning were investigated. Almost all high Schmid factor (SF>0.35) grains were twinned. A surprising high twin frequency of 82% in middle SF (0.35>=SF>=0.15) grains was observed, which was unexpected since the middle SF grains were believed to be unfavorable for twinning. The twin frequency in low SF (SF<0.15) grains was slightly increased from 2D to 3D EBSD due to the small volume of twins. The shear stress maintained a high level and was homogeneously distributed in high SF grains, facilitating twin nucleation and growth. The shear stress was distributed heterogeneously within the middle SF grains, and twins were nucleated within areas with positive shear stress. The shear stress in low SF grains was not favorable for twinning and twins occurred in the vicinity of stress accumulation. Twinning activities in the same grain varied on different layers. It was attributed to the stress fluctuation derived from grain environment changes.

**Keywords:** Mg alloy, 3D EBSD, micro-mechanics, Schmid factor, crystal plasticity modeling

## 1. Introduction

Magnesium is the lightest metallic structural material with a density of 1.738 g/cm$^3$, about 2/3 of aluminum and a quarter of iron. Due to the high critical resolved shear stress (CRSS) of non-basal slip, basal slip and tensile twinning are the dominant deformation modes at room temperature [1-3]. Tensile twinning plays an important role in accommodating the c-axis strain while basal slip can only accommodate the strain parallel to the basal plane. Another major difference is the polarity of twinning that tensile twinning is favored under tension stress along the c-axis or compression along a-axis [4]. As a result, grain orientation or texture has significant effects on twinning activities which can be estimated with Schmid's law [5]. However, the macroscopic non-Schmid twinning behavior was reported by many researchers that low Schmid factor (SF) twin variants were formed [6-9]. One explanation was the local stress which deviated from the applied stress. Also dislocation pile-ups [10, 11] and twins [12] at grain boundaries caused a sharp increase in local stress. The low SF twin variants were nucleated to release the local stress and maintain the compatibility of the material [13, 14].

Another possible reason was the shear strain accommodation. The low SF twin variants could be formed if they required less strain accommodation compared to the high SF ones [6, 15].

While the above efforts shed light on most twinning events, the appearance of some non-SF twins could not be explained by any of these theories. This was attributed to the limitation of 2D EBSD: most previous studies of twinning behaviors were based on 2D EBSD data collected from the surface layer of the sample [7, 11, 16-18]. This data contained only one section of the sample and was thus unable to reveal the real twinning occurrence across the entire 3D volume of grains. For example, grains with high twinning SF did not show twins in a 2D EBSD after deformation. However, it was not clear whether the grains had no twins across the entire volume, or just no twins appeared on this 2D section layer. Besides, there were some low SF twins formed at the boundaries which showed poor geometrical compatibility with neighboring grains [19, 20]. Were these non-SF twins initially nucleated at the grain boundaries within this layer or nucleated in bottom/upper parts of the scanned 2D EBSD layer with different grain neighbors?

To answer the above questions, 3D characterization techniques have been developed in the past decades. Three dimensional X-ray diffraction (3D XRD) is a non-destructive method to investigate the in situ deformation and recrystallization process. By using high energy X-ray, the 3D XRD method can spatially resolve the local crystallographic orientation, grain shapes, and strain states. Aydıner et al. [21] reported the evolution of stress in parent grain and twins during in situ compression. The effects of parent orientation and grain neighbor on the resolved shear stress (RSS) for twinning were investigated using synchrotron diffraction [22, 23]. However, 3D XRD can only index large grains or twins (>20 μm) with obvious diffraction patterns, leading to a relatively low resolution. Another main approach to rebuild the 3D structure of materials is serial sectioning, for example 3D EBSD in Ga Focused Ion Beam–Scanning Electron Microscope (FIB-SEM) systems and recently developed Plasma-FIB (PFIB) systems [24]. It was reported that the morphology of twins varied with SF values [25], and macroscopic non-SF twinning occurred at the low misorientation boundary. Ventura et al. [26] combined high resolution 3D EBSD and micro-tensile testing to study the twinning behaviors in pure Mg. They proposed that basal slip favored twin nucleation with a loading tilted away from the c-axis while pyramidal slip triggered twinning under c-axis loading. With a detailed layer to layer analysis of 3D EBSD data, Paramatmuni et al. [27] found that the stored energy density was a key factor to identify twin nucleation sites. Nevertheless, the examined volume by FIB-EBSD is very small due to its low milling rate, especially in the thickness direction with only tens of microns. Although the removal rate of PFIB in theory could be up to 15,000× faster than a typical Ga + FIB [28], the obtained 3D volume in practice could not be dramatically increased, like $350 \times 350 \times 235$ μm$^3$ in ref [27] and $500 \times 500 \times 100$ μm$^3$ in ref [29]. Therefore, PFIB still cannot effectively address the challenges when multi-scale microstructure features need to be investigated simultaneously in detail within one sample (i.e., investigating fine sized twins in coarse grains requires both large sampling area and small step size).

A number of theoretical and numerical frameworks have been developed to investigate the role of local stress distribution, dislocation activation, and grain neighbors on the nucleation and formation of tensile twins in Mg. For example, full-field crystal plasticity modeling and atomistic simulation are widely used. Different lenticular and irregular twin shapes were observed along twinning shear and lateral directions [30], respectively, which is related to the differences in the mobility of the edge and screw components of twinning dislocations. Gong et al. [31] found that the interaction of basal slip and twins would introduce jogs and basal stacking faults in the matrix, and prismatic <a> dislocations in the twin. Liu et al. [32-34] combined a phase field model with dislocation density crystal plasticity model to study the stress distribution and dislocation density within parent and twins.

In this work, 3D EBSD with a large volume and full-field crystal plasticity modeling are combined to investigate the twining mechanisms of a Mg alloy. This alloy with an initial weak basal texture provides numerous grains in various orientations, both favorable and unfavorable for twinning. Different from Ga or Plasma FIB-assisted 3D EBSD techniques, repetitive mechanical polishing is applied for the serial sectioning process which enables EBSD characterization with a large examined volume of 1500 × 1000 × 600 μm$^3$ and fine step size of 0.5 μm. The EBSD data shows that tensile twinning activities are underestimated in 2D EBSD. Significant increases in the twinning frequency are found when 2D grains in different layers are correlated. While the SF is related to the distribution of the RSS, it does not directly affect the twin nucleation process. The RSS is accountable to the formation of twins, and it varies from layer to layer because of different grain neighbors. These findings are expected to exhibit a statistical view of twinning in Mg alloys and bring new insights into twinning mechanisms from the aspects of grain orientation, grain neighborhood and RSS.

## 2. Methodology

2.1. Experimental procedure

A commercial WE43 alloy supplied by Luxfer MEL Technologies Ltd, as detailed in [35], was used in this study. The rectangular cubes with the dimension of 5(RD) × 5(RD) ×10(ED) mm were cut from the extruded bar (RD and ED indicate radial direction and extrusion direction, respectively), as shown in Fig. 1(a). Solution treatment at 525 °C for 1 h was carried out to dissolve secondary phases and optimize the grain size for twinning. This sample was then slightly grounded to remove the surface oxidation. Uniaxial compression was applied along one of the RD to a stress of 150 MPa, approximately the yield strength of this material [11]. This stress restricted rapid twin propagation so that the twin nucleation sites could be identified.

After that, serial sectioning and EBSD characterization were applied on the RD-RD plane of the compressed sample, as shown in Fig. 1(b). In the sectioning part, the sample was grounded using grit 4000 SiC foil and polished with 0.25 μm oil-based diamond suspension, followed by final OPS polishing.

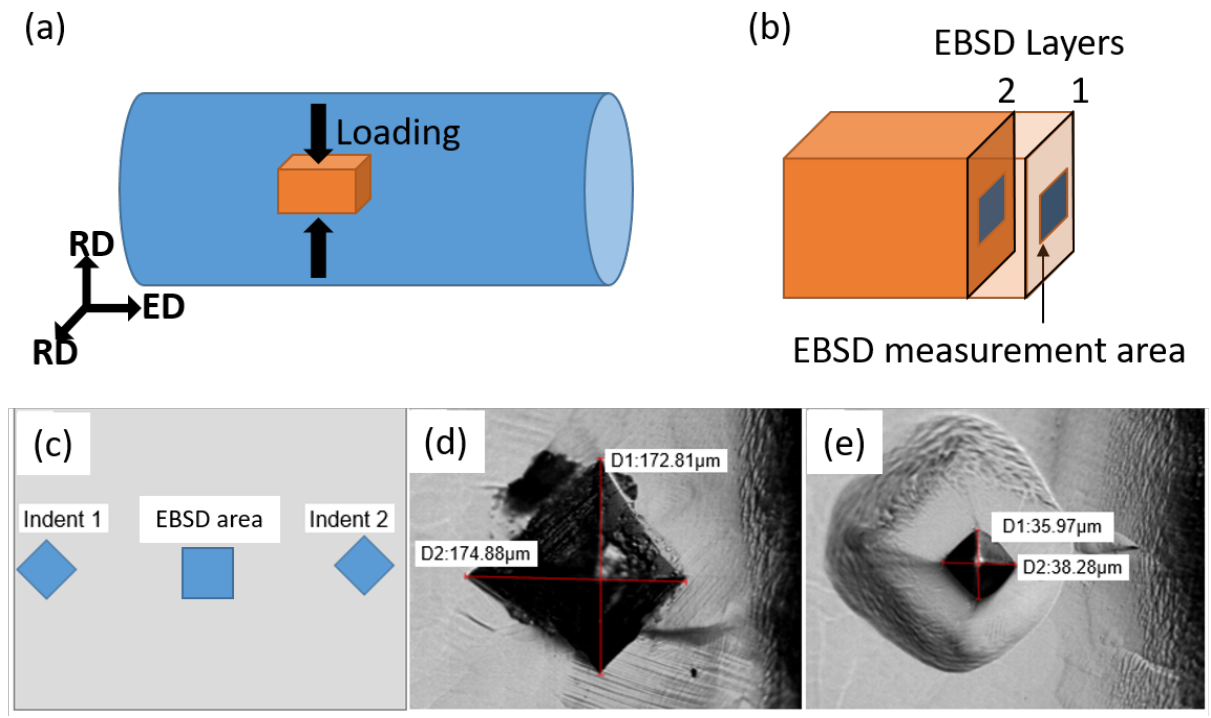

Fig. 1. (a) Schematic of the compression sample, (b) 3D EBSD via serial sectioning, (c) positions of indents and EBSD area, (d) and (e) indents before and after grinding and polishing for the 10th layer scanning, respectively.

For measurement of material removal, two micro-hardness indents were introduced at the edge of the samples before grinding and polishing, far away from the EBSD sampling area in the center, as shown in Fig. 1(c). For example, Fig. 1(d) shows one indent introduced after completing the EBSD scan of the 9th layer and before the grinding and polishing for the 10th layer scanning. The depth (h) of indent before and after grinding and polishing can be calculated using the following equation according to the geometry of the Vickers Hardness tester (average length of the diagonal D = (D1 + D2)/2, angle of indenter α = 136˚).

$$h = \frac{D}{2\sqrt{2}\tan(\alpha/2)}$$

After calculation, h is around 1/7 of D. Based on the measured diagonal length of the indents before and after grinding and polishing in Fig. 1(d-e), the sectioned depth is around 20 μm. This removal control strategy was used in the whole serial sectioning process and the spacing between every two EBSD layers is 20 ± 3 μm.

The EBSD characterization was carried out on a FEI Nova 450 field emission gun SEM equipped with an Oxford Instruments NordlysNano EBSD detector. Serial sectioning was conducted parallel to the Z-axis (extrusion direction), forming an examined volume of 1500 × 1000 × 600 μm³ with a section thickness of 20±3 μm resulting in 30 (X–Y) slices.

Table 1. Tensile twin variants in Mg alloys.

| Twin variant | Twinning system | Rotation axis |
| --- | --- | --- |
| T1 | $(1\bar{1}01)[\bar{1}120]$ | $[11\bar{2}0]$ |
| T4 | $(1\bar{1}0\bar{1})[\bar{1}10\bar{2}]$ | $[\bar{1}\bar{1}20]$ |
| T2 | $(10\bar{1}\bar{1})[\bar{1}01\bar{2}]$ | $[1\bar{2}10]$ |
| T3 | $(10\bar{1}1)[\bar{1}012]$ | $[\bar{1}2\bar{1}0]$ |
| T5 | $(01\bar{1}1)[0\bar{1}12]$ | $[\bar{2}110]$ |
| T6 | $(0\bar{1}11)[01\bar{1}2]$ | $[2\bar{1}\bar{1}0]$ |

The post analysis of the EBSD data was carried out with the MTEX package in Matlab [36]. This is necessary as the polarity of twinning mechanism is not considered in commercial EBSD software. Twin variants are indexed by the minimum deviation angle approach: (1) calculate the potential six variant orientations by the misorientation of 86 degree around the $<11\bar{2}0>$ axis, (2) calculate the misorientation angle between real twins orientation and ideal twin variant orientation, (3) index the twin variant with lowest misorientation angle which should be below 5 degree. The six tensile twin variants shown in Table 1 are categorized into three groups, T1&T4, T2&T3, and T5&T6, based on their rotation axis. For, example, twin variant T1 with $[11\bar{2}0]$ rotation axis and T4 with $[\bar{1}\bar{1}20]$ belong to the same twin variant group and they typically have comparable macroscopic SF values. For the sake of brevity, all macroscopic SF afterwards is termed as SF.

## 2.2. Full-field crystal plasticity simulations

In order to reveal more quantitative insights into the complex micromechanics of the WE43 alloy, *i.e.* local stress distributions, strain partitioning, slip system activities, and their roles in the determination of tensile twin nucleation, full-field dislocation density-based crystal plasticity simulations were carried out on the basis of the 3D EBSD microstructure mapping [32-34, 37].

The total deformation gradient $\mathbf{F}$ is decomposed into an elastic part $\mathbf{F}_e$ and a plastic part $\mathbf{F}_p$, as $\mathbf{F} = \mathbf{F}_e \mathbf{F}_p$. The evolution of $\mathbf{F}_p$ is related to the plastic velocity gradient $\mathbf{L}_p$ according to the flowing rule $\dot{\mathbf{F}}_p = \mathbf{L}_p \mathbf{F}_p$. $\mathbf{L}_p$ is given by the superposition of shear on all slip systems, as

$$\mathbf{L}_p = \sum_{\alpha=1}^{N_s} \dot{\gamma}_s^{\alpha} \, \mathbf{m}_s^{\alpha} \otimes \mathbf{n}_s^{\alpha},$$

where $\dot{\gamma}_s^{\alpha}$ describes the shear rate on slip system $\alpha$, and $\mathbf{m}_s^{\alpha}$ and $\mathbf{n}_s^{\alpha}$ indicate the slip direction and the slip plane normal of the $N_s$ slip systems. It is worth noting that the plastic deformation is assumed to be only contributed by dislocation slip in the current simulations, since the current work focuses on the nucleation of tensile and not twin growth.

The shear rate of mobile dislocations on a slip system is described by the Orowan equation as [38]:

$$\dot{\gamma} = \rho_\mathrm{m} b v_0 \exp\left[-\frac{Q_\mathrm{a}}{k_\mathrm{B} T}\left\{1-\left(\frac{|\tau_\mathrm{eff}|}{\tau_\mathrm{P}}\right)^p\right\}^q\right]\mathrm{sign}(\tau),$$

where $\rho_\mathrm{m}$ is the mobile dislocation density, $b$ is the magnitude of the Burgers vector, and $v_0$ is the reference dislocation glide velocity, $Q_\mathrm{a}$ is the activation energy for dislocation glide to overcome the obstacles, $k_\mathrm{B}$ is the Boltzmann constant, $T$ is temperature, $\tau_\mathrm{P}$ is the Peierls stress, and $p$ and $q$ are fitting parameters. $\tau_\mathrm{eff}$ is the effective resolved shear stress as the driving force for dislocation slip, and is calculated as the resolved stress reduced by the passing stress.

The evolution of the mobile dislocation density is determined by dislocation multiplication, annihilation, and dipole formation, *i.e.*

$$\dot{\rho}_\mathrm{m} = \frac{|\dot{\gamma}|}{b\Lambda} - \frac{2\breve{d}}{b}\rho_\mathrm{m}|\dot{\gamma}| - \frac{2\hat{d}}{b}\rho_\mathrm{m}|\dot{\gamma}|.$$

$\hat{d}$ indicates the distance below which two dislocations can form a dipole, and $\breve{d}$ is the distance between two dislocations below which they annihilate. $\Lambda$ is the dislocation mean free path, describing the strain hardening behavior.

The balance relation for linear momentum is described as $\mathrm{Div}\,\mathbf{P} = \mathbf{0}$, where $\mathbf{P}$ is the first Piola-Kirchhoff stress, and can be calculated from the second Piola-Kirchhoff stress $\mathbf{S}$. $\mathbf{S}$ is calculated as $\mathbf{S} = \mathbb{C}\mathbf{E}_\mathrm{e}$, where $\mathbb{C}$ is the stiffness tensor and $\mathbf{E}_\mathrm{e}$ is the elastic Green-Lagrange strain. $\mathbf{E}_\mathrm{e}$ is given by $\mathbf{E}_\mathrm{e} = \frac{1}{2}(\mathbf{F}_\mathrm{e}^\mathrm{T}\mathbf{F}_\mathrm{e} - \mathbf{I})$.

### 2.3. Simulation setup

For the full-field mechanical simulations, the model was directly created using the full 3D EBSD orientation data. Since the current work focused on the investigation of tensile twin nucleation rather than twin growth, the experimental analysis was based on the deformed microstructure at a relatively small strain of 1%. It was reasonable to assume that the grain morphology and the orientation of parent grains only had very minor changes after just 1% strain. However, the thin twins within the parent grain from the deformed EBSD data should be removed to create the initial high-fidelity simulation model. To this end, the orientation map at 1% strain was cleaned using the TSL OIM software [39, 40], *i.e.* the orientation within thin twins was assigned with the orientation from the parent grain.

In the current work, basal <a> ({0001}<11$\bar{2}$0>), prismatic <a> ({10$\bar{1}$0}<11$\bar{2}$0>), and pyramidal <c+a> ({10$\bar{1}$2}<$\bar{1}$011>) slip systems are considered to accommodate the plastic deformation. Monotonic compression loading under periodic boundary conditions was applied to the representative volume element, at a strain rate of $0.1 \text{s}^{-1}$. The material parameters for the constitutive model were calibrated by fitting the simulated and experimental stress-strain curves, as shown in Fig. 2d. The full set of material parameters is listed in Table 2.

Table 2. Material parameters for the constitutive model.

| Crystal plasticity model | $b$ (m) | $v_0$ (ms$^{-1}$) | $Q_a$ (J) | $\tau_P$ (Pa) | p | q |
|---|---|---|---|---|---|---|
| Basal | $3.20 \times 10^{-10}$ | $1.0 \times 10^{-5}$ | $7.0 \times 10^{-20}$ | $2.0 \times 10^{7}$ | 1.0 | 1.0 |
| Prismatic | $3.20 \times 10^{-10}$ | $1.0 \times 10^{-5}$ | $7.0 \times 10^{-20}$ | $9.5 \times 10^{7}$ | 1.0 | 1.0 |
| Pyramidal | $6.11 \times 10^{-10}$ | $1.0 \times 10^{-5}$ | $7.0 \times 10^{-20}$ | $1.6 \times 10^{8}$ | 1.0 | 1.0 |
| Elastic constants | $C_{11}$ (Pa) | $C_{33}$ (Pa) | $C_{44}$ (Pa) | $C_{12}$ (Pa) | $C_{13}$ (Pa) | |
| | $5.93 \times 10^{10}$ | $6.15 \times 10^{10}$ | $1.64 \times 10^{10}$ | $2.57 \times 10^{10}$ | $2.14 \times 10^{10}$ | |

## 3. Results

3.1. General information of the deformed sample

The initial microstructure and texture of the examined alloy are shown in Fig. 2. A twin-free microstructure with equiaxed grains is observed in the EBSD orientation map. In the (0001) pole figure, a weak fiber texture with the c-axis of most grains aligned perpendicular to ED can be found, in agreement with Ref. [11]. The average grain size is 60 μm, as shown in Fig. 2(c). Fig. 2(d) shows the true stress-strain curves from both simulation and measured tensile tests. The alloy starts to yield at 150 MPa (about 1% strain) which is also the stress where the test was interrupted for the investigated 3D EBSD sample.

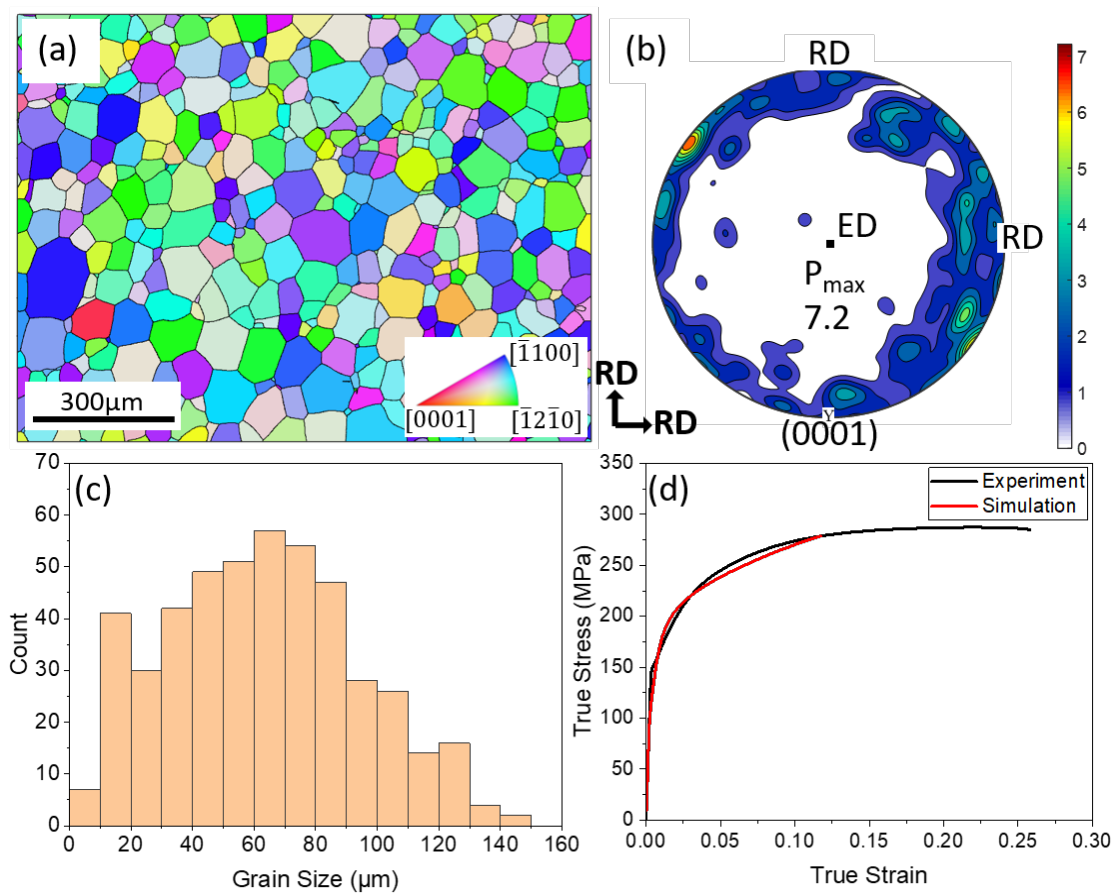

Fig. 2. Grain information of the WE43 alloy: (a) IPF map, (b) basal pole figure, (c) grain size distribution, (d) true stress-strain curve.

The 3D sectioning of the required volume resulted in 30 layers of the microstructure, which are included in Fig. S1 of the supplementary material for the sake of brevity. A random layer, layer 21, is picked to present the grain information of the sample. All the grains on layer 21 are also tracked in other layers by matching the grain orientations and positions. In Fig. 3(a), a total number of 341 grains are observed on layer 21. About one quarter of the grains have lamellar twins. The texture after compression is very similar to the initial texture due to the low volume fraction of twins at 1% strain. It should be mentioned that the same specimen reference with compression direction (CD) and radial direction aligned parallel to X and Y axis, respectively, is used for all EBSD maps afterwards. The SF map for tensile twinning is shown in Fig. 3(c). It is noteworthy that the maximum SF among six tensile twin variants is used for the SF map. The grain boundary map in Fig. 3(d) shows that more than 95% of the twins are tensile twins, which are the main focus of this work.

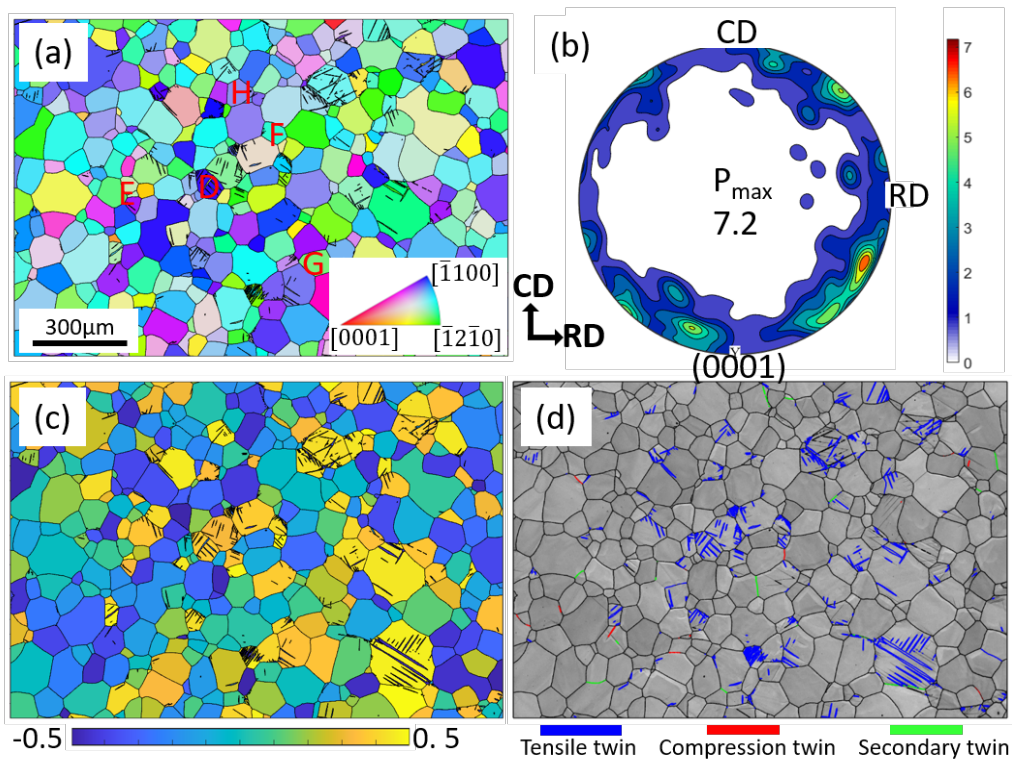

Fig. 3. EBSD maps of the compressed sample on layer 21: (a) IPF map, (b) basal pole figure, (c) SF map for tensile twinning, (d) grain boundary map.

### 3.2. Effect of Schmid Factor (SF) on twinning behavior

Since the SF plays an important role in twinning, it is of interest to study the twinning behavior of grains with different SF values for tensile twinning. All grains on layer 21 are categorized into three groups: high tensile twinning SF grains (0.5>=SF>0.35), middle SF grains (0.35>=SF>=0.15) and low SF grains (-0.5=<SF<0.15). The number of grains in the different SF groups are 61, 61, and 219, respectively. Two thirds of the grains have low SF values which means the texture is not favorable for twinning. Detailed SF maps showing these different SF groups can be found in Fig. 4. The figure shows that most twins are found in high SF grains while twinning is almost negligible in low SF grains. Besides, the number of twins per grain and diameter of twins are obviously higher in high SF grains than in other groups. A statistical evaluation of twin frequency is shown in Table 3. It is not surprising that the twin frequency on layer 21 decreases from 85% to 11% as the SF decreases which shows the dependency of twinning on SF. When the grains on layer 21 are tracked in 3D EBSD data, the twin frequencies of all SF groups are increased, which is detailed later in section 3.4. Almost all the grains (98%) with high SF are twinned in 3D EBSD. Moreover, the twin frequency of middle SF grains jumps from 54% to 82%, indicating that twinning activity in middle SF grains is underestimated in 2D EBSD. The twin frequency increased only slightly from 11% to 16% in low SF grains. This statistic analysis confirmed that 3D EBSD reveals more reliable grain and twin information which might be missing in 2D characterization.

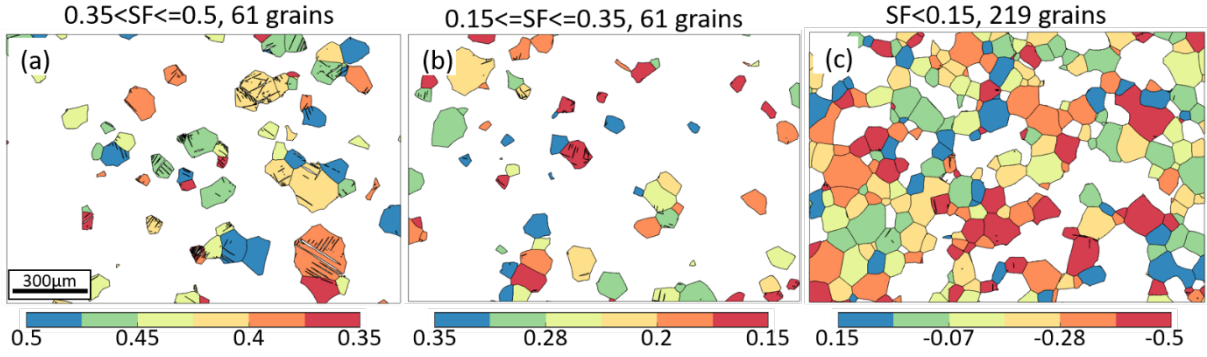

Fig. 4. Schmid factor (SF) maps of the different grain groups on layer 21.

Table 3. Twinning frequency of grain groups with different Schmid factor (SF).

| Tensile Twinning SF value | Number of grains | Twin frequency on layer 21 | Twin frequency in all layers |
| --- | --- | --- | --- |
| 0.35-0.5 | 61 | 85% | 98% |
| 0.15-0.35 | 61 | 54% | 82% |
| -0.5-0.15 | 219 | 11% | 16% |
| All | 341 | 25% | 36% |

3.3. Effects of SF on resolved shear stress (RSS)

The drawback of using the SF for twin prediction is that it assumes that the local stress in the individual grains is the same to the applied macroscopic stress. However, this assumption fails due to the complicated co-deformation of grains in polycrystalline materials. For example, dislocation pile-ups and twins in a neighboring grain can cause sharp peaks in local stress. Besides, the effect of the grain boundary is not considered while it is the preferred site for twin nucleation. In fact, the local stress is the direct driving force for twin nucleation. To investigate the effects of local stress on twin morphology and variant selection, crystal plasticity modeling is carried out based on the 3D EBSD data.

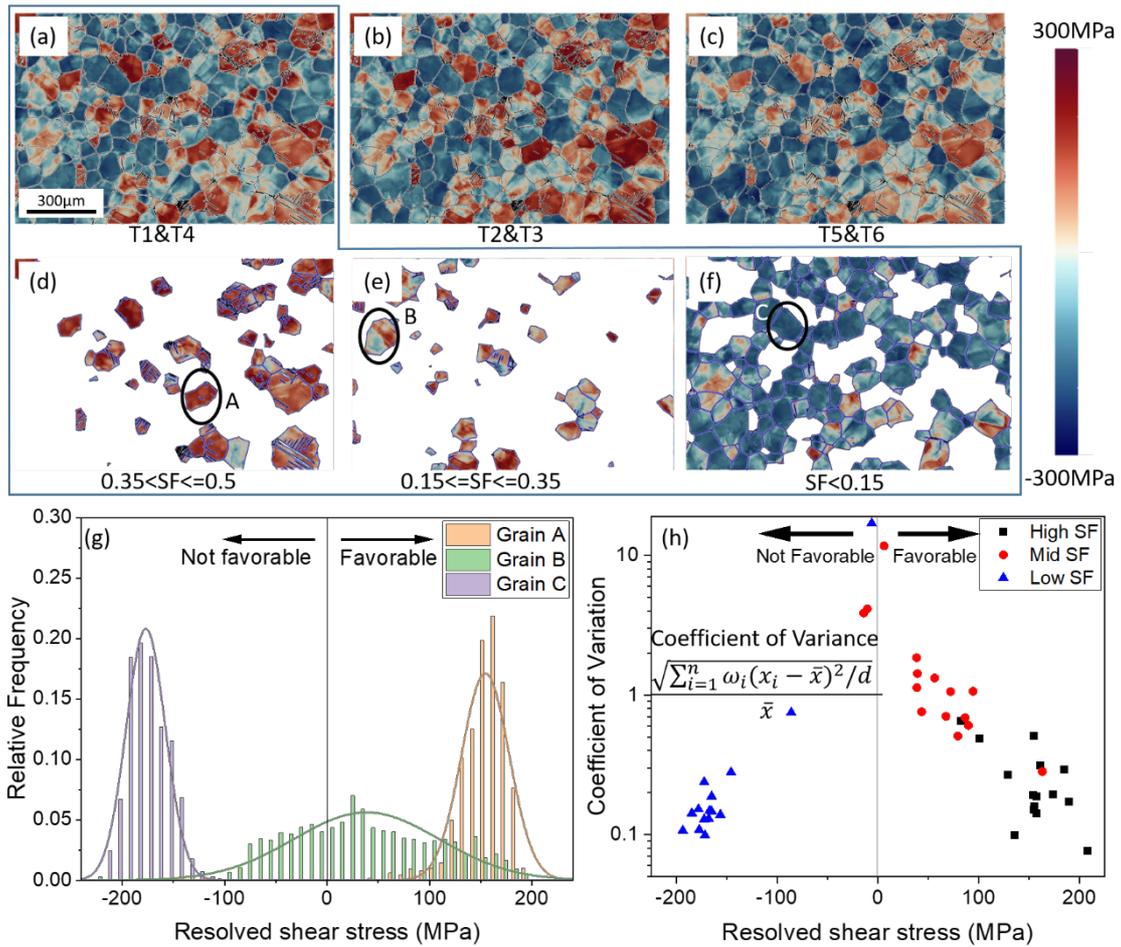

Fig. 5. RSS maps of layer 21 for: (a-c) the different twin variant groups, (d-f) different SF grains, (g) RSS distribution of three representative grains as shown in (d-f), (h) statistical coefficient variance of different SF grains.

The RSSs of layer 21 for different twin variants are shown in Fig. 5 while those of other layers can be found in Fig. S2. Since the twin variants in the same group, like T1 and T4, share an equivalent twinning plane and direction (Table 1), the distributions of RSS for them are similar. Thus, only three RSS maps for different twin variant groups are shown here which are overlaid with grain boundary maps. An inhomogeneous distribution of the RSS is observed, some grains (marked in blue) show negative RSS while others (red) have high positive RSS. Fig. 5(d-f) illustrates the effect of the SF on the RSS, taking the RSS for T1&T4 group as an example. The first thing to note is that high SF grains are normally colored in dark red which means they have very high RSS. On the contrary, most of the grains with low SF are blue with negative RSS. The RSS for middle SF grains show a combination of positive and negative RSS. In other words, heterogeneity of intragranular RSS in middle SF grain is observed while the distributions of RSS in high and low SF grains are more uniform. This tendency is confirmed in Fig. 5(g) where three representative grains from these three SF groups, grain A, B and C circled in Fig. 5(d-f), are picked to show their RSS distribution. Grain A (or C) shows intense peaks at highly positive (or negative) RSS. The RSS of grain B varies in a wide range. It should also be mentioned that the RSS for different twin variants in the high SF grains maintain a high level despite of minor fluctuations, for example grain A with a high RSS for T1&T4 also shows relatively high RSS for T2&T3 and T5&T6 in Fig. 5(a-c). A statistical evaluation of the coefficient variance of the RSS from 15 randomly picked grains in each SF group (shown in Fig. S3) is shown in Fig. 5(h). The coefficient variance of both high and low SF grains are mostly below 0.2, indicating a homogeneous distribution of the RSS in the grain. Meanwhile, the variance of middle SF grains are around 1, much higher than the former two groups. This different distribution of RSS affects the twin size which will be discussed later.

3.4. 3D characterization of twins in different SF grains

As mentioned above, distinctive twinning activities are observed in different SF groups. This seems to be related to the RSS distribution. To have a more precise view, one or two representative grains in each SF group are picked, and the corresponding 3D structures are studied in detail in the following.

3.4.1. High tensile twinning SF grains

On layer 21, a twinned and an untwinned grain with high SF, grain D and E, respectively, are shown in Fig. 6. In grain D, some large twins are found to span through the parent, as shown in Fig. 6(a). The twins are aligned parallel to each other. On layer 18, the same grain is found to have a large dimension and more twins are observed. The calculated orientations of six tensile twinning variants are shown in Fig. 6(c). By matching the real and potential twin orientations, these twins are indexed as T1, T3, and T4 according to the definition in Table 1. The variety of twins is attributed to the minor difference in SF value. Grain E in the center of Fig. 6(d) has a maximum tensile twinning SF of 0.46, but it does not have any twins on layer 21. However, this grain actually has several twins on layer 17 and the size of grain E is much larger than it appears on layer 21. While the highest SF twin variants T1 and T4 (SF=0.44, 0.46) are the dominant twin variants, some low SF variant T6 (SF=0.06) occur at the conjunction of different grains. The whole morphology of these two grains on different layers are illustrated in the form of 3D EBSD and the corresponding RSS maps are shown in Fig. 6 (g-h). It is clear that large twins existed on almost all layers of high SF grains. In addition, the RSS is evenly distributed through the matrix and maintains a high level on most layers. The deviation of grain E between 2D and 3D EBSD confirms that 3D characterization is indispensable to explore the real environment information around the investigated grains.

3.4.2. Middle tensile twinning SF grains

The twinning behaviors in middle SF grains are shown in Fig. 7. Compared to Fig. 6(a), fewer twins are observed in grain F, and the twins are much smaller in Fig. 7(a). This seems reasonable as the parent grain has a lower SF (SF=0.18), thus is less favorable for twinning. But this is not always the case for 3D EBSD characterization. Some large twins and different variants are observed on layer 16 despite the SF being relatively low, as shown in Fig. 7(b) and (c). An untwinned grain G with middle SF on layer 21 is shown in Fig. 7(d). However, tensile twins are observed on other layers as observed for grain E. While high twinning activities are observed in almost all layers of the high SF grains, the middle SF grains show various degrees of twinning (in amount and size) on different layers, as shown in Fig. 7(g) and (h). In terms of RSS distribution, grain F shows high RSS on layer 14-16 which exhibit advanced twinning activities, and low RSS are observed on layer 17-21 in agreement with the small twins. This link between RSS and twinning activities is also confirmed in grain G with large twins and high RSS on layer 23 and 24.

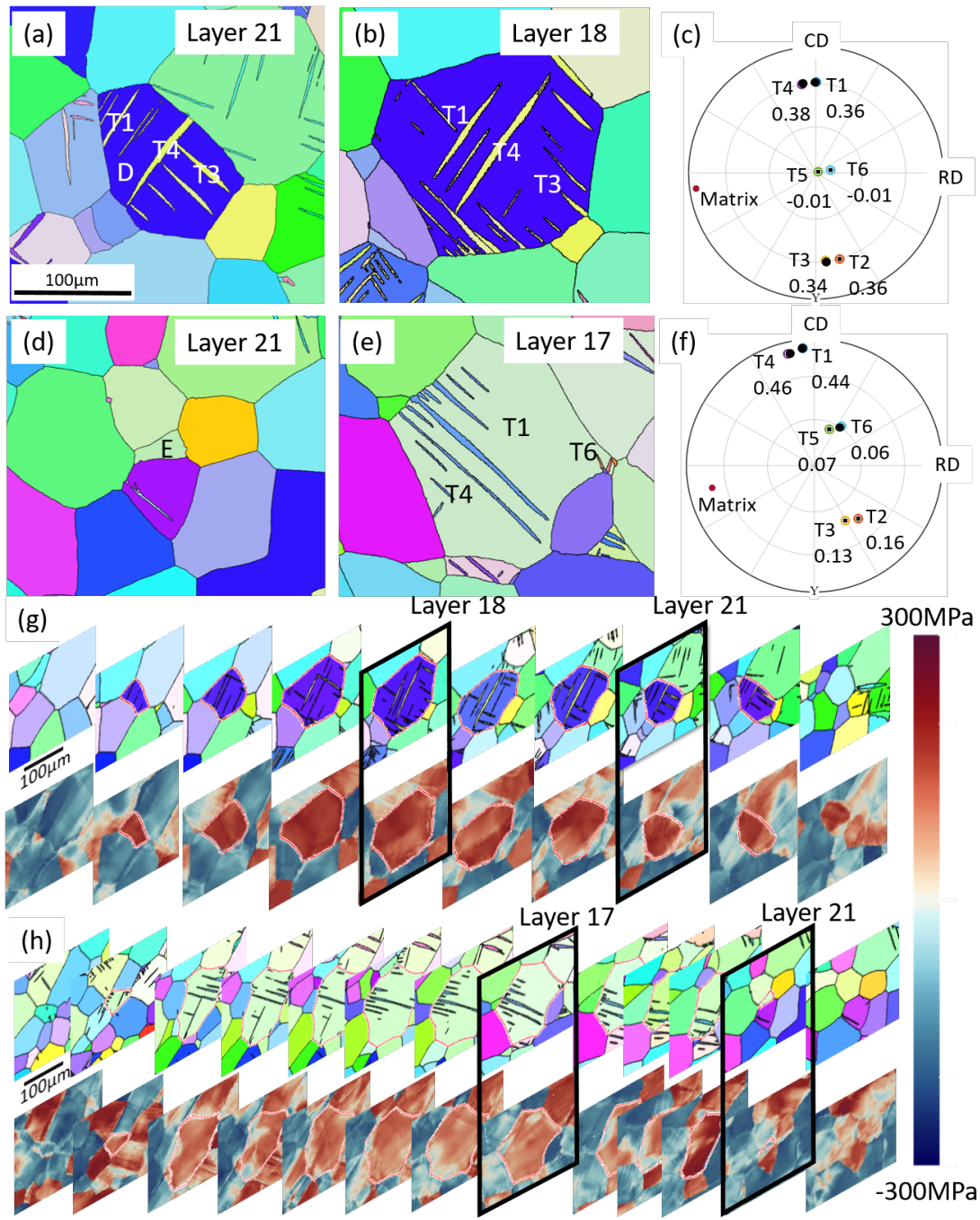

Fig. 6. EBSD maps of high SF grains: (a-b) twinned and (d-e) untwinned grain in different layers, (c) and (f) corresponding orientation of matrix and twin variants, (g) and (h) 3D EBSD and RSS of the grains.

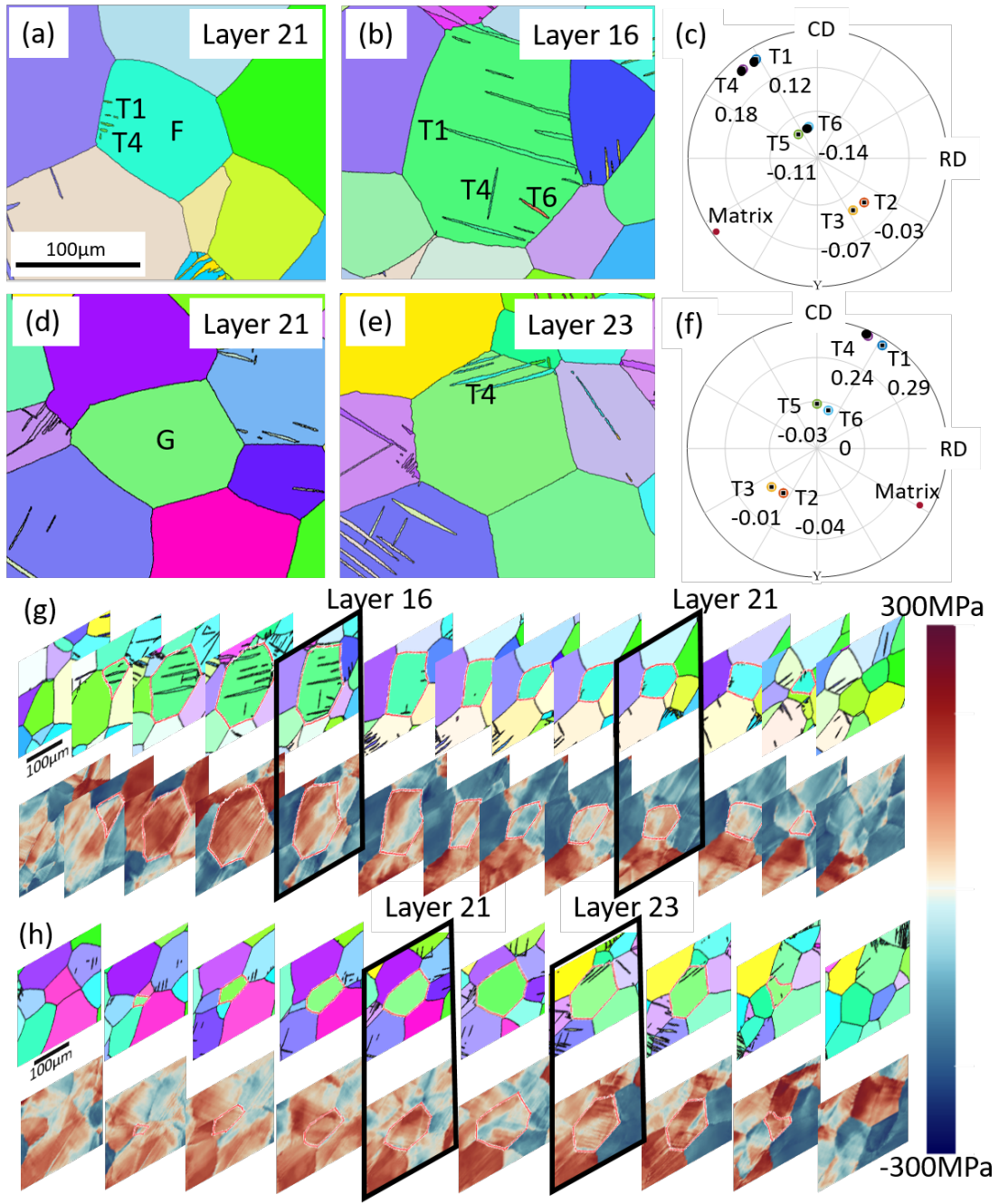

Fig. 7. EBSD maps of middle SF grains: (a-b) twinned and (d-e) untwinned grain in different layers, (c) and (f) corresponding orientation of matrix and twin variants, (g) and (h) 3D EBSD and RSS of the grains.

3.4.3. Low tensile twinning SF grain

In terms of twinning in low SF grains, one example of a twinned grain is shown in Fig. 8. Only two tiny twins are found, as shown in Fig. 8(a). The grain boundary map in Fig. 8(c) shows that these twins are nucleated from the right side boundary of the grain. The twins are only observed on layer 21, Fig. 8(e). Based on the basal pole map in Fig. 8(d), these twins are indexed as variant T4 with a SF of 0. In this case, the calculated shear stress on variant T4 is 0 and theoretically twins should not occur. However, a RSS peak is observed on layer 21 which matches the location of twins in Fig. 8(e), although the RSS is generally negative in most layers. The low SF grains are not favorable for twinning and twin growth is hindered.

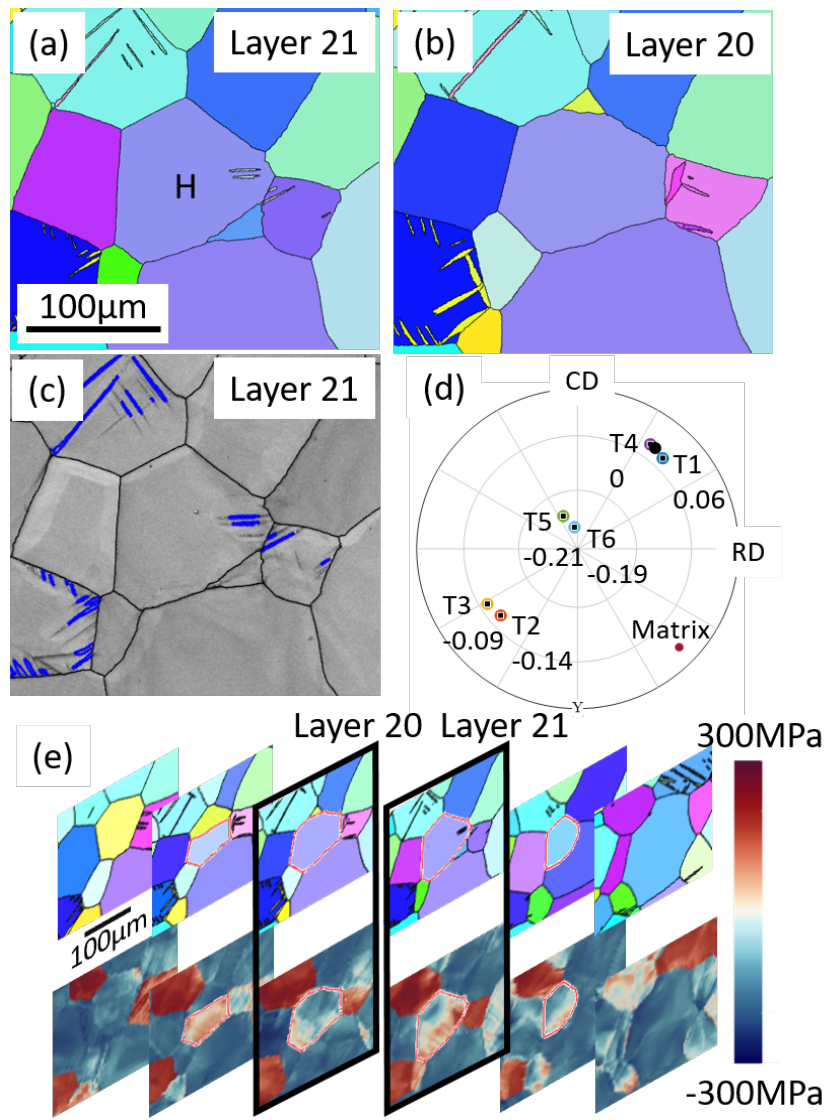

Fig. 8. EBSD maps of middle SF grains: (a-b) twinned and grain in different layers, (c) band contrast and boundary map, (d) corresponding orientation of matrix and twin variants, (e) 3D EBSD and RSS of the grains.

### 3.5. Effect of grain neighbors on twin nucleation

The strain accommodation from grain environment (e.g., geometric compatibility factor, m') have been used in explaining non-SF twinning activities when these twins have a high m' [19, 20]. However, in some cases, tension twin variants with low SF and low m' were still observed. One possible explanation is these tension twins are nucleated above or below the examined surface of 2D EBSD where the twinned parent grain has different neighboring grains. However, it has been a challenge to get all the neighboring grain information in 2D EBSD. The current 3D EBSD allows us to investigate the changes of grain neighbor relationships. As grains are in different depths of the materials, some new grains may be found and some may disappear on certain layers. Fig. 9(a) shows the orientation map of layer 20. Comparing to layer 21, 42 out of 320 grains (marked with red boundaries) on layer 20 are different from those exist on layer 21. These different grains definitely introduce new boundaries and affect the strain accommodation. The example in Fig. 9(b) shows the orientation map of grain D on different layers. On layer 21, grain D has six neighboring grains, as marked in the figure. By tracking the same grain in the 3D EBSD data, another three grains come into contact with grain D on layer 20, introducing different grain pair relationships. The total number of neighbors of grain D through all layers is 18, much higher than that observed on layer 21. Fig. 9(c) shows the statistical data of neighboring grain number from 25 randomly picked grains. The number of neighboring grains is dependent on the grain size as large grains tend to span more layers and have more neighboring grains. The average number of neighboring grains on layer 21 and all layers for these grains are 7 and 17, respectively. This means that more than half of the neighboring grain information is missing in 2D EBSD.

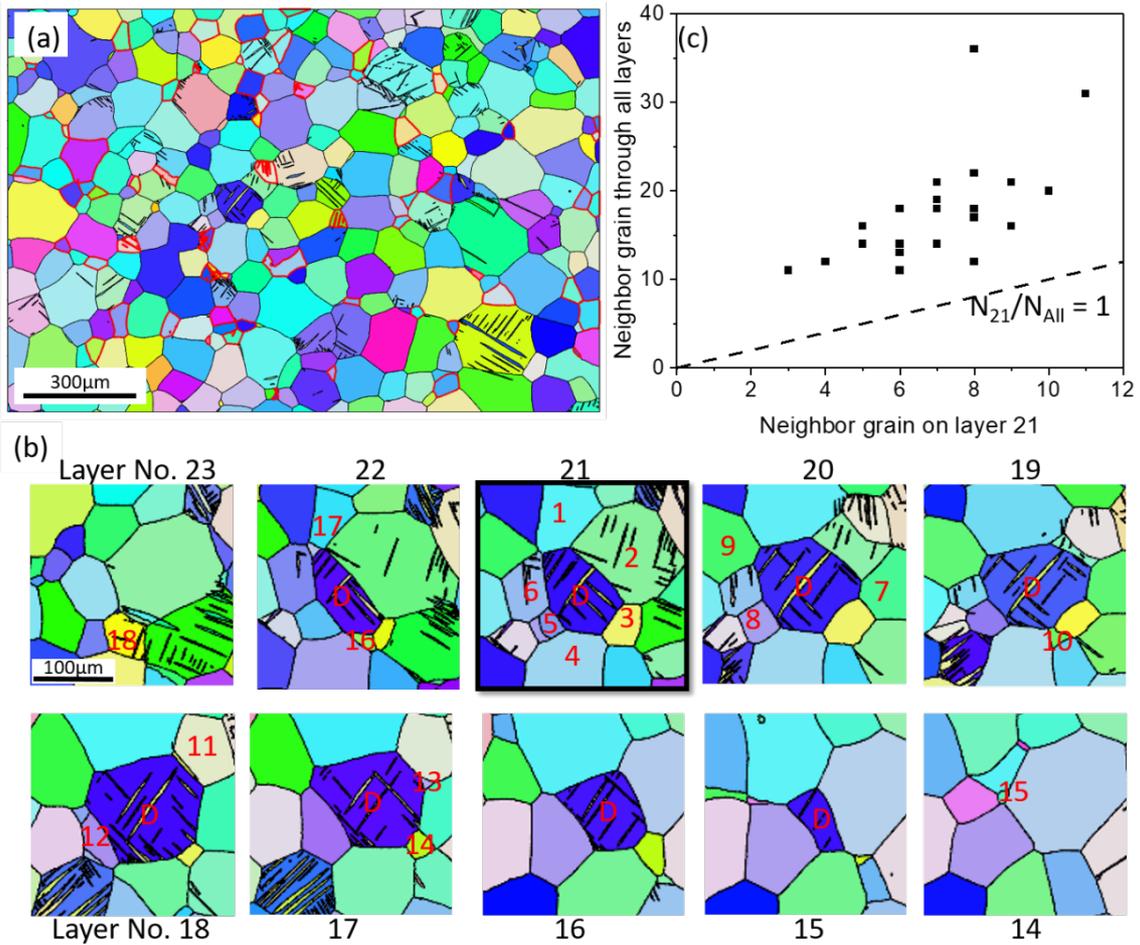

Fig. 9. (a) EBSD map of layer 20, (b) neighbors of grain D on different layers, (c) number of neighboring grain on layer 21 ($N_{21}$) versus all layers ($N_{All}$).

As mentioned in section 3.4, the observed twinning activity in the same grain can vary from layer to layer. This is attributed to the fluctuation of RSS caused by the different grain environments. How neighboring grains can affect the RSS is shown by the following example. Grain J, K, and L are adjacent grains on layer 21, as shown in Fig. 10(a). Grain J and K are twinned while grain L is twin-free. Their basal poles are aligned perpendicular to the compression load (Y axis), indicating that they are all favorable for twinning, Fig. 10(g). The corresponding RSS map in Fig. 10(d) shows a relatively low RSS region at their adjacent boundaries. In Fig. 10(h), the RSS of grain K along the path 5-1-2 near adjacent boundaries stays negative. When it comes to layer 20, grain M, oriented unfavorable for twinning, is found to separate the former three grains. Some new tiny twins are observed in grain L and J, and the twins are larger in grain K, as shown in Fig. 10(b). Grain L has a negative RSS due to its hard orientation. Compared to the RSS on layer 21, the RSS in grain J, K, and L are obviously increased on layer 20, Fig. 10(e). The average RSS of grain K is doubled from layer 21 to 20, as shown in Fig. 10(h). A similar phenomenon is also observed on layer 23 where the RSS of grain K is increased along the path 1-2 at the boundary with hard orientated grain N.

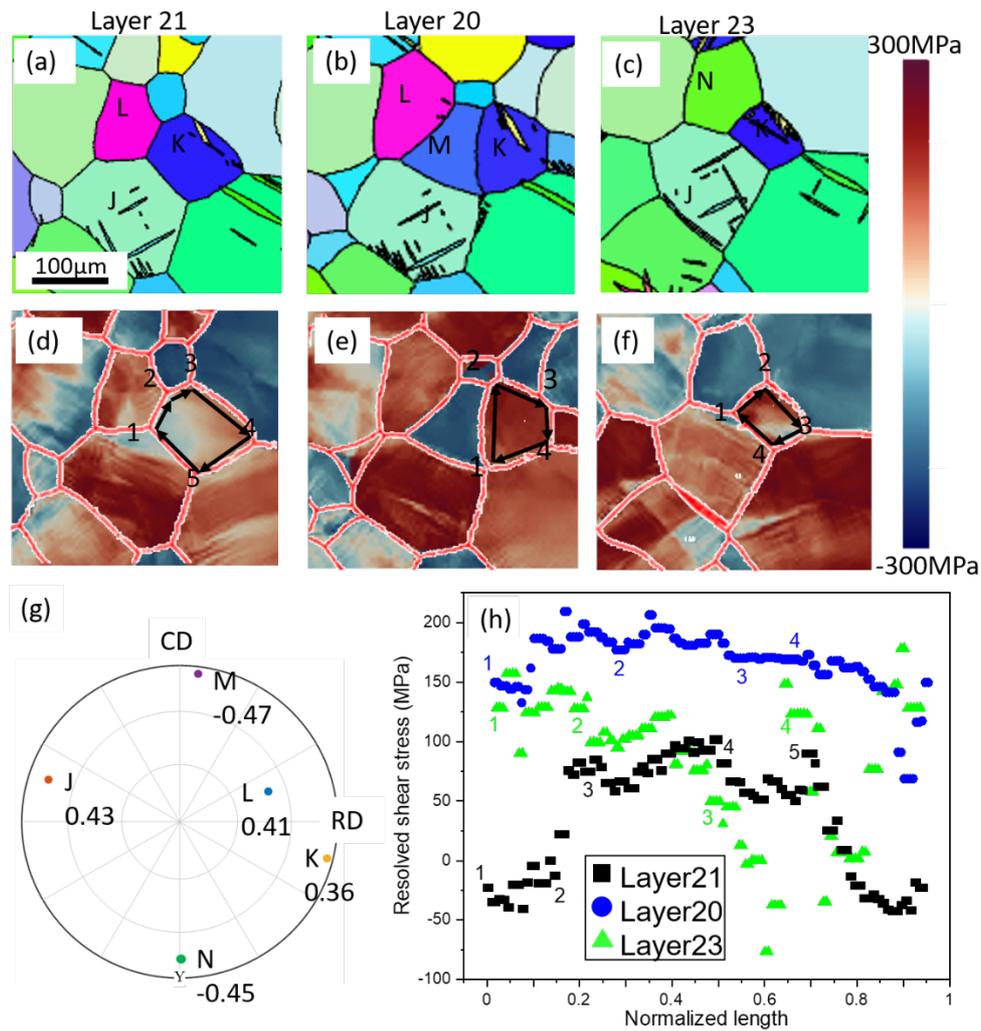

Fig. 10. Effect of neighboring grains on the RSS: (a-c) EBSD maps of grains in different layers, (d-f) corresponding RSS maps, (g) grain orientations and their twinning SF, (h) RSS along boundary areas.

## 4. Discussion

It is well known that basal slip and tensile twinning are the dominant deformation modes in Mg alloys at room temperature. Because the CRSS ratio of basal slip to tensile twinning in Mg single crystals was reported to be in the range of 1:2-4 [41, 42], it appears that basal slip is more likely to be activated than tensile twinning. One of our findings is that tensile twinning is underestimated in 2D EBSD compared to 3D EBSD. This is shown in Table 3, i.e. the overall twin frequency increases by 50% from layer 21 (25%) to all layers in 3D EBSD (36%). Since 2D EBSD can only exhibit one section of the grain, it can happen that twins occur outside of this section, as shown in Fig. 6(d) and Fig. 7(d). Consequently, the contribution of twinning to the deformation process of Mg alloy is underestimated. A surprising high twin frequency of 82% is observed in middle SF grains, much higher than reported in the work of other researchers [7, 43] using conventional 2D EBSD. It should be mentioned that the sample is only deformed up to the yield stress of 150 MPa. According to the equation $\sigma_{twin} = \sigma \times SF$ [5], the calculated shear stress for tensile twinning in middle SF grains ranges from 22.5 to 52.5 MPa. This shear stress is insufficient to stimulate so many twins (82% twin frequency) in middle SF grains as it is lower than the reported CRSS for tensile twinning (85 MPa) in WE43 [44] and other Mg alloys [45, 46]. In addition, the actual twin frequency is expected to be higher than 16% in low SF grains because the twins are too small to be indexed. These profuse twins formed at yielding will continue to grow as well as new twins will be formed with further loading, indicating that strain accommodation by twinning is underestimated in 2D EBSD.

Twinning is not solely controlled by the SF as twins are found in low and middle SF grains. The RSS is the direct driving force for twin nucleation and growth which depends on the SF and other factors [47]. The emerging technique 3D XRD is capable of measuring the stress in the twinned grain [21, 22]. However, only the grain-average stress instead of local stresses is obtained in 3D XRD while the local stresses can reach a high level due to inhomogeneous deformation. In this work, crystal plasticity modeling is used to simulate the RSS. The links between twinning behaviors, SF, and RSS are established. As is observed in Fig. 5, the RSS is high and homogeneously distributed in high SF grains while in low SF grains it is generally negative, in agreement with Abdolvand's work [48] showing that favorably oriented grains have higher RSS than others. High RSS provides the stress necessary for twins to grow at the twin tip [49], and thus twins in high SF grains show higher growth rates and consume large volume of the matrix. In middle SF grains, the distribution of RSS is heterogeneous and both positive and negative RSS can be observed in the same grain. Twins can only be nucleated and grow in the local areas with positive RSS, leading to a smaller diameter and lower frequency of twins than in high SF grains. This explains why the twin frequency in middle SF grains increases so dramatically from 2D to 3D EBSD. Some small twins which are not shown on a single 2D layer are more likely to be observed by taking more measurements at different depths of the grain in 3D EBSD. The majority of low SF grains shows negative RSS, unfavorable for twinning. Only a few tiny twins are found in grains with high RSS peaks. As a result, EBSD measurements can rarely capture the twinning area. The loss of twins from 2D to 3D EBSD increases with decreasing the twin size.

The grain environment plays a critical role in twinning behaviors because the more deformed neighboring grains can exert back stresses on the less deformed grain, stimulating twin nucleation. The effects of neighboring grains are widely reported in 2D EBSD [11, 12, 50, 51]. With respects to 3D investigation of the grain environment on twinning, Bieler et al. [22] found slip transfer from soft neighboring grains would stimulate the low SF twin variants. Abdolvand et al. [52] investigated the strong effect of the grain neighbors and found the stress in grains surrounded by hard oriented neighbors decreased without twinning. The current work is the first time, to the knowledge of the authors, to statistically show the difference in grain neighbors between 2D and 3D EBSD, Fig. 9. It confirms that more than half of the grain neighbor information is missing in 2D EBSD. Similar to the study by Paramatmuni [27], a detailed layer by layer analysis of the RSS with different grain neighbors is shown in Fig. 10. Since strain can be easily accommodated by twinning between the favorably oriented grains, the local RSS is low in the adjacent boundary region, as shown in Fig. 10(h). However, the strain accommodation can be hindered by a hard oriented grain on another layer like shown in Fig. 10(b). In this case, the RSSs of the former three grains are increased, for example the RSS of grain K is doubled. This definitely promotes twinning activities.

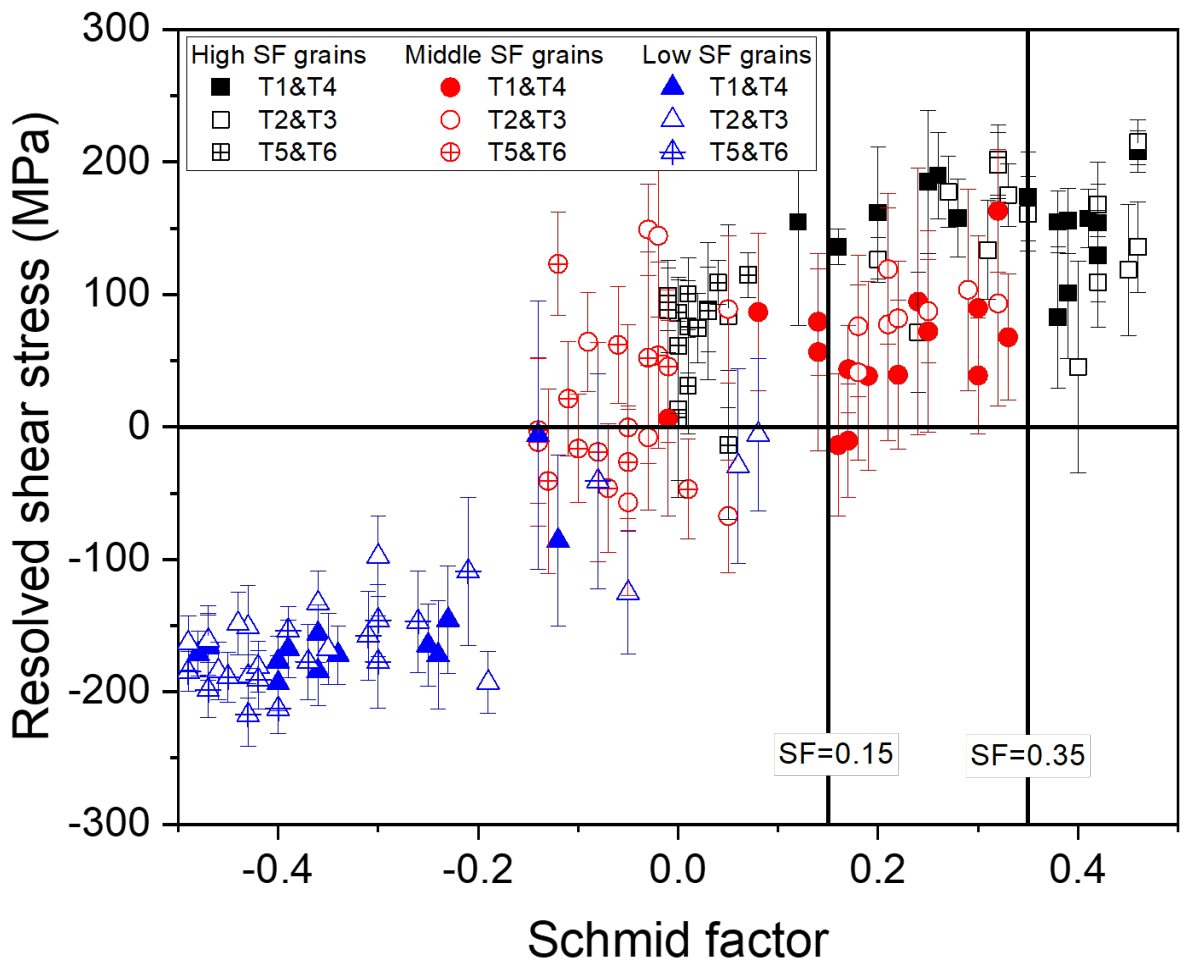

Fig. 11. Distribution of the average RSS of grains for different twin variants.

Due to the high twining activity, most of the non-SF twin variants occur in high SF grains. This is because the intense twinning activities in high SF grains can enhance the local stress, for example stress can accumulate at twin-twin intersections. As shown in Fig. 5(a-c), although the RSSs of high SF grains for three twin variant groups are different, they all stay at a high level which could be sufficient to stimulate non-SF twins. The grains from Fig. 5(h) are used to investigate non-SF twin variants and RSS. Fig. 11 shows the average RSS and corresponding SF for different twin variants. It is shown that high SF grains may have middle or low SF values for some twin variants, as is the case for T5&6 in grain E in Fig. 6(e). However, their RSSs mostly stay above 100 MPa. The average RSS of all twin variants in high SF grains is 121 MPa, about three times higher than that of middle SF grains, 44 MPa. In contrast, low SF grains tend to show low or negative RSS for twinning with an average of -150 MPa. As a result, only a few twins are observed in localized area with RSS peak owing to the effect of neighboring grains, Fig. 8.

## 5. Conclusions

Tensile twinning in Mg alloy was statistically investigated by 3D EBSD and crystal plasticity modeling. This is the first time that the spatial tensile twinning activities in a large bulk Mg sample are investigated. This large dataset will be used for machine learning in future work. Various twinning activities were observed on different layers as well as in different SF grains. This phenomenon was interpreted with respect to the RSS. The main conclusions drawn are:

1. Tensile twinning is underestimated as the twin frequency increased by 50% from 25% for 2D EBSD to 36% in 3D EBSD characterization. High twinning activities in middle SF grains are already observed at yield stress despite the texture not being favorable for twinning.

2. High SF grains are likely to show a high twin frequency and large twins compared to the low SF counterpart. Twinning activity in the same grain varies on different layers.

3. The RSS is the driving force for twinning. High SF grains tend to show high and homogeneously distributed RSS which promotes twin nucleation and growth. On the contrary, low SF grains have negative RSS that hinders twin nucleation. As a result, twins are restricted in areas of high local RSS.

4. The change of grain environment affects the RSS distribution. The strain cannot be easily accommodated between soft and hard orientated grains which leads to an increase in the RSS, and subsequently enhances the twinning activity.

5. Most of the non-SF twin variants are found in high SF grains owing to the high level of RSS.

**Acknowledgements**

This work was supported by the UKRI Future Leaders Fellowship, MR/T019123/2. CL and FR acknowledge financial support by the Deutsche Forschungsgemeinschaft (DFG) within projects A01 and C01 of the Collaborative Research Center (SFB) 1394 "Structural and Chemical Atomic Complexity - from defect phase diagrams to material properties", project ID 409476157. CZ and JD acknowledge the financial supports of the National Natural Science Foundation of China (Grant No. 52071208).